\documentclass[a4paper]{INTERSPEECH2024}

\usepackage{graphicx}
\usepackage{amssymb,amsmath,bm}
\usepackage{textcomp}
\usepackage{url}
\usepackage[utf8]{inputenc}
\usepackage{CJKutf8}
\usepackage{multirow}
\usepackage{comment}
\usepackage{subfigure}

\newcolumntype{P}[1]{>{\centering\arraybackslash}p{#1}}

\ninept
\interspeechcameraready

\title{Noise-Robust Voice Conversion by Conditional Denoising Training\\ Using Latent Variables of Recording Quality and Environment}

\name[affiliation={1}]{Takuto}{Igarashi}
\name[affiliation={1}]{Yuki}{Saito}
\name[affiliation={1}]{Kentaro}{Seki}
\name[affiliation={1,2}]{Shinnosuke}{Takamichi}
\name[affiliation={3}]{Ryuichi}{Yamamoto}
\name[affiliation={3}]{\\Kentaro}{Tachibana}
\name[affiliation={1}]{Hiroshi}{Saruwatari}

\address{
$^1$The University of Tokyo, Japan,
$^2$Keio University, Japan,
$^3$LY Corp., Japan.
}
\email{
takuto0228@g.ecc.u-tokyo.ac.jp
}

\begin{document}
\setlength{\abovedisplayskip}{10pt} 
\setlength{\belowdisplayskip}{10pt} 
\setlength\floatsep{12pt} 
\setlength\intextsep{10pt} 
\setlength\textfloatsep{10pt} 
\setlength{\dbltextfloatsep}{10pt} 
\setlength{\dblfloatsep}{10pt}

\maketitle

\begin{abstract}

We propose noise-robust voice conversion (VC) which takes into account the recording quality and environment of noisy source speech.
Conventional denoising training improves the noise robustness of a VC model by learning noisy-to-clean VC process.
However, the naturalness of the converted speech is limited when the noise of the source speech is unseen during the training. To this end, our proposed training conditions a VC model on two latent variables representing the recording quality and environment of the source speech.
These latent variables are derived from deep neural networks pre-trained on recording quality assessment and acoustic scene classification and calculated in an utterance-wise or frame-wise manner. As a result, the trained VC model can explicitly learn information about speech degradation during the training.
Objective and subjective evaluations show that our training improves the quality of the converted speech compared to the conventional training.
\end{abstract}

\section{Introduction}\label{sect:intro}
Voice conversion~(VC) is a technology that converts a source speaker's timbre to that of a target speaker while preserving the linguistic content.
VC can be widely applied in the real world, such as in movie dubbing~\cite{Mukhneri20}, personalized text-to-speech~\cite{Kain98}, and speaking assistance~\cite{nakamura2012speaking}.
As a result of advancements in deep learning, deep neural network~(DNN)-based VC methods have significantly improved the quality of converted speech~\cite{zhao2020voice}.

Typical DNN-based VC methods train a VC model with a large multi-speaker corpus containing high-quality speech from a variety of speakers.
However, actual speech samples recorded in the real world are often degraded by various factors, such as background noise and recording channels.
Huang et al.~\cite{huang21} empirically demonstrated that the recording-quality mismatch of input speech between the training and inference (i.e., clean and noisy) significantly deteriorates VC performance.

Denoising training (DT)~\cite{huang2022toward} is one promising approach to achieving noise-robust DNN-based VC.
In DT, a VC model is trained using pseudo-noisy (i.e., artificially degraded) speech, with the aim of implicitly denoising input speech during the VC process.
Although this training mitigates the distortion of converted speech caused by noisy input speech in inference, the naturalness of the converted speech is still limited when the degradation factor of input speech is unseen during the training.
One primary reason is that the trained VC model does not explicitly learn information about speech degradation, such as noise characteristics (e.g., stationary or non-stationary) and noise levels, and does not guarantee generalization performance to various degradation factors.

In this study, we propose {\it conditional DT (CDT),} an improved version of conventional~(i.e., unconditional) DT, to improve the noise robustness of VC towards unseen degradation.
CDT conditions a DNN-based VC model on two latent variables regarding the degradation of input speech: recording quality and environment.
These latent variables are derived from deep neural networks pre-trained on recording quality assessment and acoustic scene classification and calculated in an utterance-wise or frame-wise manner.
As a result, the trained VC model can explicitly learn information about speech degradation during the training.
We present the CDT framework using NISQA~\cite{mittag21nisqa} and PaSST~\cite{koutini2021efficient} as representative models to extract latent variables of the recording quality and environment, respectively.
In the experimental evaluation, we validate the effectiveness of CDT using S2VC~\cite{lin21s2vc} as the baseline VC model following the conventional DT framework~\cite{huang2022toward}.
Our contributions are summarized as follows.
\begin{itemize}
    \item We propose \textit{CDT} that can improve the noise-robustness of DNN-based VC models by explicitly conditioning the model on speech degradation information.
    \item We present two conditioning strategies with an utterance-wise and frame-wise manner so that the trained VC model can take into account not only global but also time-variant characteristics of degradation in input speech.
    \item From the evaluation results, we show that conditioning the VC model on frame-wise latent variables of recording environment is essential for improving the naturalness of the converted speech in the noisy-to-clean VC scenario.
\end{itemize}

\section{Conventional DT-based noise-robust VC}
\subsection{DT algorithm}\label{subsect:dt}
DT is an end-to-end learning method for noisy-to-clean VC, in which data augmentations are utilized to train a noise-robust VC model.
In the learning process, pseudo-noisy speech is artificially generated from clean speech by mixing various environmental noises with random signal-to-noise ratios~(SNRs) and fed into the VC model as input.
The training objective function is computed by comparing the ground-truth clean speech and converted speech, i.e., the VC model's output.
Typically, the L1 or L2 loss between mel-spectrograms of ground-truth clean speech and converted speech is used as the objective function.
Thus, DT can be interpreted as a learning method in which the VC model acts as a denoising autoencoder~\cite{huang2022toward}.

Huang et al.~\cite{huang2022toward} demonstrated that the noise robustness of VC models trained by DT improved when the VC models were based on an autoencoding process such as AdaIN-VC~\cite{chou2019one} and S2VC~\cite{lin21s2vc}, and S2VC with DT was the most effective for VC.

\subsection{Limitations}\label{subsect:problem}
Another approach to achieving noise-robust VC without modifying the model structure can be the concatenation of a pre-trained speech enhancement model with an existing any-to-any VC model.
However, this approach tends to limit the VC performance compared to an end-to-end approach such as DT.
This is because the artifacts caused by speech enhancement to suppress unseen noises negatively affects the downstream VC task~\cite{zhang2021denoispeech}.

Although DT is an end-to-end approach, the naturalness of the converted speech is still limited when the noise of the input speech is unseen during the training.
One of the main reasons is that the trained VC model does not explicitly learn information about speech degradation, such as noise characteristics and noise levels, and does not guarantee generalization to various degradation factors.

\begin{figure}[t]
    \centering
    \subfigure[NISQA]{\includegraphics[width=0.4\linewidth, clip]{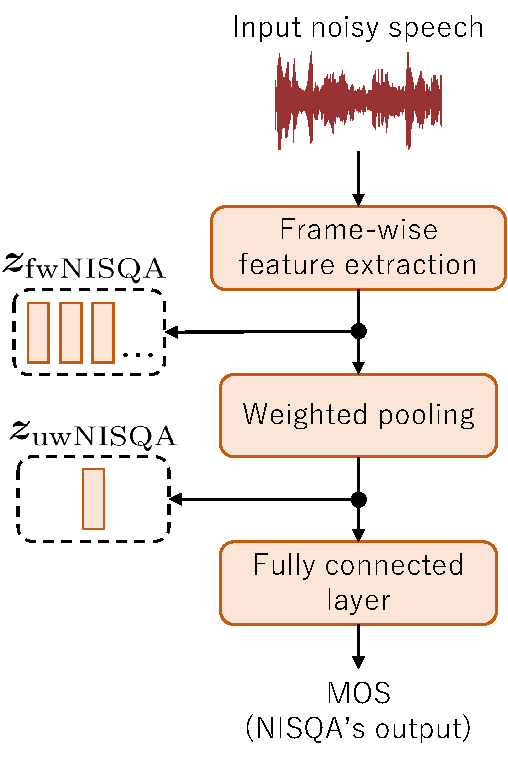}}\hspace{5mm}
    \subfigure[PaSST]{\includegraphics[width=0.3\linewidth, clip]{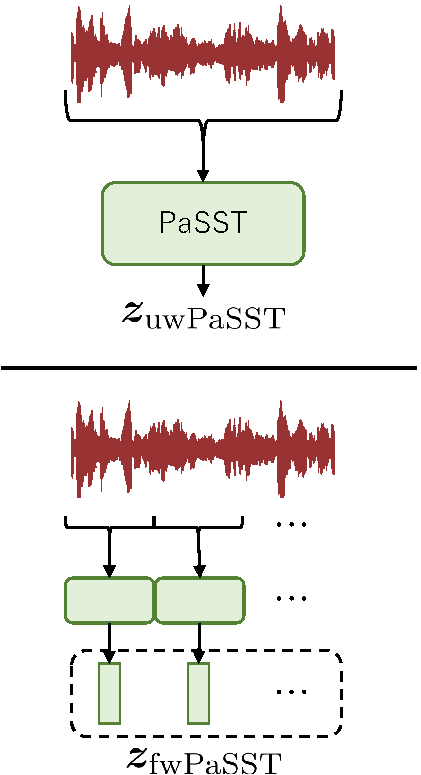}}
    \vspace{-4mm}
    \caption{Latent variables extracted by NISQA and PaSST in an utterance-wise or frame-wise manner.}
    \label{fig:extraction}
\end{figure}

\section{Proposed CDT-based noise-robust VC}
\subsection{Motivation}\label{subsect:motivation}
One possible solution to the problem presented in Section~\ref{subsect:problem} is to enable the DNN-based VC model to explicitly learn information about speech degradation.
To this end, we propose CDT, which conditions the model on two latent variables regarding the degradation of input speech: recording quality and environment.
For practicality, our CDT algorithms assume that only the source speech samples are degraded by noise.
This is because it is more challenging for VC users, who are not always experts in speech technology or high-quality speech recording, to record their own clean source speech than it is to collect target speakers' clean speech.
Nevertheless, we believe that our algorithm can be extended to VC where both the source and target speech are degraded.

\subsection{CDT algorithm}
Let $\bm x^{\mathrm{c}}$ be any clean speech in the training dataset.
The loss function $\mathcal L$ of our CDT is defined as follows:
\begin{align}
    \bm x^{\mathrm{s}} &= \bm x^{\mathrm{c}} + \bm n,\label{eq:xs}\\ 
    \bm x^{\mathrm{t}} &= \bm x^{\mathrm{c}},\\
    \bm z_{\mathrm{SSL}}^{\mathrm{s}} &= f_{\mathrm{SSL}}\left(\bm x^{\mathrm{s}} \right),\\
    \bm z_{\mathrm{rqa}}^{\mathrm{s}} &= f_{\mathrm{rqa}}\left(\bm x^{\mathrm{s}} \right),\\
    \bm z_{\mathrm{asc}}^{\mathrm{s}} &= f_{\mathrm{asc}}\left(\bm x^{\mathrm{s}} \right),\\
    \bm z_{\mathrm{SSL}}^{\mathrm{t}} &= f_{\mathrm{SSL}}\left(\bm x^{\mathrm{t}} \right),
\end{align}
\begin{equation}
    \mathcal{L} = \left | f_{\theta}\left(\bm z_{\mathrm{SSL}}^{\mathrm{s}}, \bm z_{\mathrm{rqa}}^{\mathrm{s}}, \bm z_{\mathrm{asc}}^{\mathrm{s}}, \bm z_{\mathrm{SSL}}^{\mathrm{t}}\right) - g_{\mathrm{mel}}(\bm x^{\mathrm{c}}) \right |, \label{eq:L}
\end{equation}
where $\bm x^{\mathrm{s}}$ and $\bm x^{\mathrm{t}}$ are the source and target speech, respectively.
The pseudo-noisy source speech $\bm x^{\mathrm{s}}$ is calculated by adding noise $\bm n$ to the clean speech $\bm x^{\mathrm{c}}$.
$g_{\mathrm{mel}}(\cdot)$ is a function which calculates the log mel-spectrogram from input speech.
$f_{\mathrm{SSL}}(\cdot)$ is a pre-trained SSL model that extracts intermediate feature representations from the source and target speech used for the VC process, i.e., $\bm z_{\mathrm{SSL}}^{\mathrm{s}}$ and $\bm z_{\mathrm{SSL}}^{\mathrm{t}}$.
The source speaker's conditional latent variables for recording quality and environment, i.e., $\bm z_{\mathrm{rqa}}^{\mathrm{s}}$ and $\bm z_{\mathrm{asc}}^{\mathrm{s}}$, are extracted by pre-trained DNNs for recording quality assessment $f_{\mathrm{rqa}}(\cdot)$ and acoustic scene classification $f_{\mathrm{asc}}(\cdot)$, respectively.
$f_{\theta}(\cdot)$ is a DNN-based model parameterized by $\theta$, which predicts the target speaker's log mel-spectrogram from the input features.
The model parameter $\theta$ is updated to minimize the log mel-spectrogram prediction error shown in Eq.~\eqref{eq:L}.

We present the CDT framework using NISQA~\cite{mittag21nisqa} and PaSST~\cite{koutini2021efficient} as representative models to extract latent variables of the recording quality and environment: $\bm z_{\mathrm{rqa}}^{\mathrm{s}}$ and $\bm z_{\mathrm{asc}}^{\mathrm{s}}$, respectively.

{\bf NISQA: }
NISQA is a method for automatically estimating recording quality scores without reference speech.
The open-source model is trained on pseudo-noisy and real-noisy speech, allowing it to robustly predict recording quality values against a variety of noises.

{\bf PaSST: }
The model trained on AudioSet~\cite{gemmeke2017audio}, which contains 527 types of tags, achieves significantly higher predictive performance compared to other models~\cite{koutini2021efficient}.

\subsection{Frame-wise conditioning}
NISQA and PaSST were originally designed to output an utterance-wise prediction: the mean opinion score (MOS) and audio tag, respectively.
However, frame-wise features can be obtained as described in the next paragraph.
The frame-wise features can be expected to represent the non-stationary characteristics of noise in the noisy source speech.

Let $(\bm z_{\mathrm{uwNISQA}}, \bm z_{\mathrm{uwPaSST}})$ and $(\bm z_{\mathrm{fwNISQA}}, \bm z_{\mathrm{fwPaSST}})$ be latent variables extracted by (NISQA, PaSST) in an utterance-wise and frame-wise manner, respectively.
Figure~\ref{fig:extraction} shows the feature extraction methods.
As shown in Figure~\ref{fig:extraction}(a), NISQA's model segments an input speech, estimates frame-wise features~$\bm z_{\mathrm{fwNISQA}}$, computes their weighted average~$\bm z_{\mathrm{uwNISQA}}$ along with the frame axis, and outputs the final prediction (MOS of input speech).
In contrast, PaSST's model outputs audio tags with no dimensions in the frame direction, as shown in the upper part of Figure~\ref{fig:extraction}(b).
However, we can also use the model to extract frame-wise audio tags by segmenting the input speech in advance~\cite{turian2022hear}.
The former extracts $\bm z_{\mathrm{uwPaSST}}$ and the latter extracts $\bm z_{\mathrm{fwPaSST}}$.

\begin{figure*}[t]
  \centering
  \includegraphics[width=0.7\linewidth, clip]{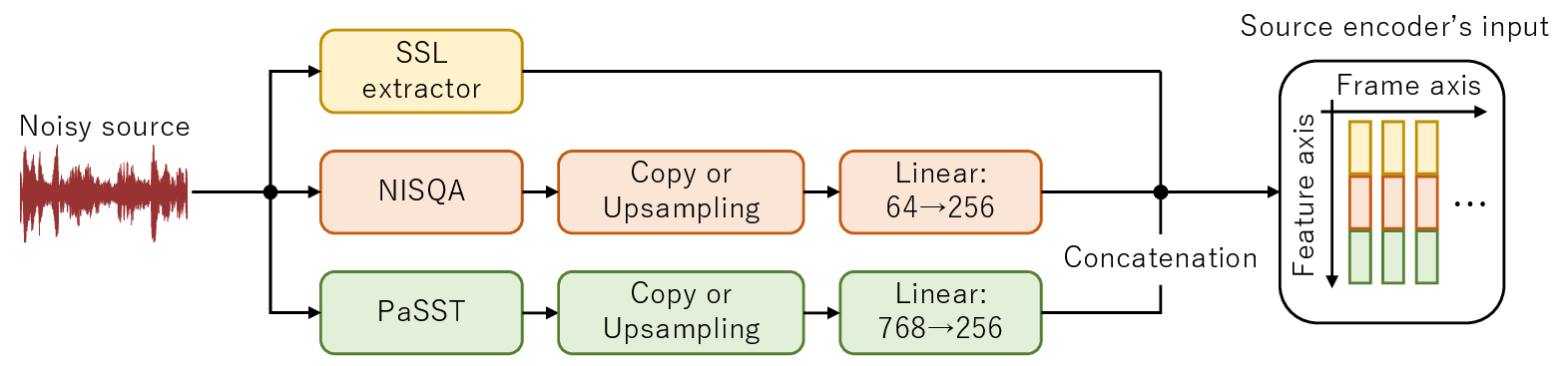}
  \vspace{-4pt}
  \caption{Conditioning the VC model on latent variables}
  \label{fig:concat}
\end{figure*}

\section{Experimental evaluation}
\subsection{Experimental conditions}
We used the parallel100 subset from Japanese Versatile Speech (JVS)~\cite{takamichi2019jvs} and downsampled all speech data to $16$~kHz.
The subset contains $22$ hours of speech data for $100$ Japanese speakers ($100$ utterances per speaker).
The numbers of speakers included in the training, validation, and evaluation data were $90$ (``jvs001'' to ``jvs086''), $4$ (``jvs087'' to ``jvs090''), and $10$ (``jvs091'' to ``jvs100''), respectively.
The $10$ test speakers excluded from the training data were used as unseen speakers for the any-to-any VC evaluation.

We created the pseudo-noisy speech dataset by adding noise~$\bm n$ to the clean speech~$\bm x^{\mathrm{c}}$ with SNR randomly sampled from a uniform distribution $U(0, 20)$ [dB].
Here, $\bm n$ was taken from the DEMAND~\cite{thiemann2013demand} noise and the WHAM!48kHz~\cite{Wichern2019WHAM} noise for the training and evaluation datasets, respectively.
Thus, the noise of speech from the evaluation dataset was unseen in the training process.
We downsampled the WHAM!48kHz to $16$~kHz.
Both contained various types of background noises.

Referring to Huang et al.'s study~\cite{huang2022toward}, we adopted S2VC~\cite{lin21s2vc} as a backbone VC model with the publicly available implementation on GitHub (robust-vc)\footnote{\url{https://github.com/cyhuang-tw/robust-vc}}.
In CDT, we conditioned the VC model on ($\bm z_{\mathrm{uwNISQA}}$ or $\bm z_{\mathrm{fwNISQA}}$) and ($\bm z_{\mathrm{uwPaSST}}$ or $\bm z_{\mathrm{ fwPaSST}}$) as shown in Figure~\ref{fig:concat}.
We unified the size of these latent variables to that of $\bm z_{\mathrm{SSL}}$, i.e., 256 $\times$ frame size in this study.
We concatenated the latent variables along with the feature axis and input them to the source encoder.
First, to unify the frame dimension of ($\bm z_{\mathrm{uwNISQA}}, \bm z_{\mathrm{uwPaSST}}$) to that of $\bm z_{\mathrm{SSL}}$, we replicated them by a factor of the frame size of $\bm z_{\mathrm{SSL}}$ along with the frame axis.
Meanwhile, we upsampled ($\bm z_{\mathrm{fwNISQA}}, \bm z_{\mathrm{fwPaSST}}$) to align the frame lengths.
Second, to unify the feature dimensions of ($\bm z_{\mathrm{uwNISQA}}, \bm z_{\mathrm{fwNISQA}}$) and ($\bm z_{\mathrm{uwPaSST}}, \bm z_{\mathrm{fwPaSST}}$), we inserted the linear projection layers from 64 to 256 dimensions and from 768 to 256 dimensions, respectively. 

In CDT, the S2VC model was trained based on the loss function shown in Eq.~\eqref{eq:L} while the parameters of the pre-trained models, $f_{\mathrm{SSL}}(\cdot)$, $f_{\mathrm{rqa}}(\cdot)$, and $f_{\mathrm{asc}}(\cdot)$, were fixed.
Specifically, $f_{\mathrm{SSL}}(\cdot)$ is a feature extractor using contrastive predictive coding~\cite{oord2018representation} as is the case of the study of Huang et al.~\cite{huang2022toward}.
The latent variable extractors, $f_{\mathrm{rqa}}(\cdot)$ and $f_{\mathrm{asc}}(\cdot)$, were based on the official implementations of NISQA\footnote{\url{https://github.com/gabrielmittag/NISQA}} and PaSST\footnote{\url{https://github.com/kkoutini/passt\_hear21}}, respectively.
In the implementation of NISQA, the frame length was $150$~ms and the frame shift was $40$~ms.
In the implementation of PaSST, the former was $160$~ms and the latter was $50$~ms.
To generate the converted speech waveform from the predicted log mel-spectrogram, we used the HiFi-GAN vocoder~\cite{kong2020hifi} trained on the JVS training data with a batch size of 16, while all the VC models were trained with a batch size of 6.
The optimizer was AdamW~\cite{loshchilov2017decoupled} with a learning rate of $5\times 10^{-5}$, and $\beta_1 = 0.9, \beta_2 = 0.999$.
Each of the VC models had approximately 33 million parameters which were randomly initialized.
The training was stopped when the validation loss converged completely and the training time was approximately 60 hours.
We trained the VC model using both the conventional DT~\cite{huang2022toward} and our CDT frameworks and evaluated their effectiveness.

\subsection{Objective evaluation}\label{subsec:objective}
We randomly selected $250$~test pairs of source and target speech samples with different speakers taken from the evaluation dataset~(10 JVS speakers).
Then, we performed VC on each pair with the VC models trained by the conventional DT and CDT, which took less than 80 ms.
Our CDT methods are denoted as ``uwNISQA-uwPaSST'', ``uwNISQA-fwPaSST'', ``fwNISQA-uwPaSST'', and ``fwNISQA-fwPaSST'', depending on whether the model was conditioned on utterance-wise or frame-wise features.

To evaluate the intelligibility of the converted speech, we used the character error rate~(CER) estimated by the automatic speech recognition~(ASR) system.
The ASR model for calculating CER is a pre-trained ReazonSpeech model available on HuggingFace\footnote{\url{https://huggingface.co/reazon-research/reazonspeech-espnet-next}}.
The smaller the CER, the more the converted speech preserves the linguistic content of the source speech, and the larger the CER, the more severely distorted the converted speech.
Thus, CER is regarded as a measure reflecting the intelligibility of the converted speech.
In contrast, to measure the speaker similarity between the converted speech and the target speaker, we used the speaker embedding cosine similarity~(SECS).
To compute SECS, we generated two fixed-dimensional embedding vectors representing the speaker identity of the converted and target speech and computed their cosine similarity.
The higher the SECS, the more similar the converted and target speech are in terms of the speaker identity.
We adopted x-vector~\cite{snyder2018xvector} extracted by using a pre-trained model of WavLM~\cite{chen2022wavlm} as the embedding vectors, which is available on HuggingFace\footnote{\url{https://huggingface.co/microsoft/wavlm-base-sv}}.

\begin{table}[t]
\centering
\caption{Average CER and SECS of the 250 samples converted by the VC models trained with conditional DT and CDT.
The method without any conditioning corresponds to conventional DT.
``uw'' means utterance-wise and ``fw'' means frame-wise.
Values in the parentheses indicate standard deviations.
}
\label{tb:objective}
\begin{tabular}{cc|cc}
\hline
\hline

    \multicolumn{2}{c|}{Method} & \multirow{2}{*}{CER [\%]} & \multirow{2}{*}{SECS} \\ 
    \cline{1-2}
    NISQA & PaSST & & \\
    \hline
    - & - & $26.3$ & $0.935$ $(\pm 0.034)$\\
    uw & uw & $27.2$ & $\bm{0.938}$ $(\pm 0.036)$\\
    uw & fw & $24.6$ & $0.935$ $(\pm 0.034)$\\
    fw & uw & $24.7$ & $0.931$ $(\pm 0.036)$\\
    fw & fw & $\bm{23.3}$ & $0.934$ $(\pm 0.033)$\\
\hline
\hline
\end{tabular}
\end{table}

Table~\ref{tb:objective} shows the average CER and SECS of the converted speech samples corresponding to the 250 test pairs.
As shown, SECS is almost constant across methods, which may be because the VC models trained by CDT received the same information on target speech as the conventional DT when only the source speech was noisy.
In contrast, CER differed between the methods.
Although NISQA and PaSST were originally designed to output utterance-wise prediction, we can see from Table~\ref{tb:objective} that conditioning the VC model on the utterance-wise features is not very effective for improving CER.
In particular, the CER of ``uwNISQA-uwPaSST'' is lower than that of the conventional DT.
This may be because the utterance-wise features consist of the exact same matrices along the frame axis.
Although they are quite large, they do not contain much useful information for VC.
Thus, they can prevent the VC model from receiving information that is essential for VC.
In contrast, conditioning the model on frame-wise features improved CER.
This implies that the frame-wise features represent the non-stationary characteristics of noise in the noisy source speech and that the VC model leverages useful information from the conditional latent variables to improve VC performance.

\subsection{Subjective evaluation}
We conducted subjective evaluations using crowdsourcing on Lancers\footnote{\url{https://www.lancers.jp/}} regarding the naturalness and speaker similarity of converted speech.
We combined every 250 converted speech samples generated in Section~\ref{subsec:objective} into a single dataset and used it as the dataset for subjective evaluation.
Thus, the evaluation dataset contained $250\times 5=1250$~converted speech samples. When evaluating their naturalness, evaluators were given a converted utterance randomly sampled from the subjective evaluation dataset.
Then, they rated the perceptual quality on a 5-point MOS scale from 1~(very bad) to 5~(very good).
When evaluating speaker similarity, evaluators were given a ground-truth target utterance and the converted utterance sampled from the JVS corpus and the subjective evaluation dataset, respectively.
Then, they answered how similar the speakers were who produced the two utterances on a MOS score ranging from 1~(completely different) to 5~(completely same).
In both cases, the listening experiment was repeated 20 times per evaluator.
There were 100 evaluators for each subjective evaluation on naturalness and speaker similarity.

\begin{table}[t]
\centering
\caption{
Naturalness and speaker similarity of the speech samples converted by conventional DT and CDT. Values in the parentheses indicate 95\% confidential intervals of the scores. {\bf Bold} scores are the highest among the five compared methods.
}
\label{tb:subjective}
\begin{tabular}{cc|cc}
\hline
\hline

    \multicolumn{2}{c|}{Method} & \multirow{2}{*}{Naturalness} & \multirow{2}{*}{Speaker similarity} \\ 
    \cline{1-2}
    NISQA & PaSST & & \\
    \hline
    - & - & $2.75$ $(\pm 0.085)$ & $2.43$ $(\pm 0.082)$ \\
    uw & uw & $2.67$ $(\pm 0.084)$ & $2.39$ $(\pm 0.082)$ \\
    uw & fw & $2.84$ $(\pm 0.087)$ & $\bm{2.50}$ $(\pm 0.082)$ \\
    fw & uw & $2.74$ $(\pm 0.086)$ & $2.43$ $(\pm 0.083)$ \\
    fw & fw & $\bm {2.85}$ $(\pm 0.086)$ & $2.47$ $(\pm 0.082)$ \\
\hline
\hline
\end{tabular}
\end{table}

\begin{table}[t]
\centering
\caption{
Preference AB test results for naturalness of converted speech for any pair of (Conventional DT, uwNISQA-fwPaSST, fwNISQA-fwPaSST). ``Con. DT'' means conventional DT, ``(uw, fw)'' means uwNISQA-fwPaSST, and ``(fw, fw)'' means fwNISQA-fwPaSST.
}
\label{tb:ab}
\begin{tabular}{c|c|c}
\hline
\hline
A vs B & Naturalness & p-value \\
\hline
Con. DT vs (uw, fw) & $0.414$ vs $\bm{0.586}$ & $< 10^{-6}$\\
Con. DT vs (fw, fw) & $0.442$ vs $\bm{0.558}$ & $< 10^{-3}$\\
(uw, fw) vs (fw, fw) & $\bm{0.514}$ vs $0.486$ & $0.38$\\
\hline
\hline
\end{tabular}
\end{table}

As the subjective evaluation results in Table~\ref{tb:subjective} show, ``uwNISQA-uwPaSST'' demonstrated the lowest VC performance in terms of both naturalness and speaker similarity as is the case of the CER results shown in Table~\ref{tb:objective}.
On the other hand, ``uwNISQA-fwPaSST'' and ``fwNISQA-fwPaSST'' outperformed the conventional DT, although there is no statistical significance between the scores.

To investigate the differences between ``Conventional DT'', ``uwNISQA-fwPaSST'', and ``fwNISQA-fwPaSST'', we further conducted preference AB tests to compare the naturalness of the converted speech for any pair of these three methods.
Fifty listeners took part in each test, and each listener evaluated 10 pairs of converted speech samples using our crowdsourcing-based evaluation platform. 
The results are shown in Table~\ref{tb:ab}.
The proposed methods~(i.e. ``uwNISQA-fwPaSST'' and ``fwNISQA-fwPaSST'') significantly outperformed the conventional DT ($p < 0.05$), but there was no significant difference between ``uwNISQA-fwPaSST'' and ``fwNISQA-fwPaSST''.
This indicates that conditioning the VC model on $\bm z_{\mathrm{fwPaSST}}$ is effective for improving the naturalness of the converted speech in the noisy-to-clean VC scenario.

\section{Discussion}
We investigated noise-robust VC from noisy source speech to clean noisy speech in this paper and demonstrated the effectiveness of CDT.
We anticipate that CDT should be applicable to VC where the target speech is also noisy.
Noisy-to-noisy VC~\cite{xie22icassp,yao23icassp}, which aims to preserve background noise of source speech during the VC process, is another possible situation in which CDT can be applied.

We used two latent variables, recording quality and environment, as the conditional features on the DT algorithm. Other variables that characterize input speech, such as speech naturalness (e.g., UTMOS)~\cite{saeki2022utmos} and reverberation (e.g., T60 estimator~\cite{choi23interspeech}), can be also introduced to our CDT.

Regarding the training objective function, we only considered the simple L1 loss between the ground-truth and generated mel-spectrograms.
We can introduce other machine learning techniques to improve the noise robustness of the trained VC model, such as adversarial training with feature decoupling~\cite{chen24nrvc}.
 
\section{Conclusion}
We proposed conditional denoising training~(CDT), which conditions a VC model on two latent variables regarding the recording quality and acoustic environment of noisy source speech.
We verified the effectiveness of four CDT methods in the cases where these two latent variables were utterance-wise or frame-wise.
The objective and subjective evaluations showed that conditioning the VC model on frame-wise features can effectively improving VC performance, while conditioning it on utterance-wise features does not necessarily improve VC performance.

In the future, we will consider VC models that take into account various degradations of input speech, including reverberation and bandwidth rejection, as well as the additive noise considered in this study.
In addition, towards real-world applications, we will extend our CDT to address noisy speech recorded by actual devices such as smartphones.

\textbf{Acknowledgements:}
This research was conducted as joint research between LY Corporation and Saruwatari-Takamichi Laboratory of The University of Tokyo, Japan. This work was supported by Research Grant S of the Tateishi Science and Technology Foundation.

   \newpage
   \ninept
   \bibliographystyle{IEEEtran}
   \bibliography{template}

\end{document}